\documentstyle[12pt,cite,epsf]{article}
\newcommand{\mframe}[1]{{#1}}
\newcommand{\ang}[1]{{\langle #1 \rangle}}
\newcommand{\kt}{\tilde{\ae}}
\newcommand{\ktt}{\tilde{\tilde{\ae}}}
\newcommand{\bktt}{\beta \ktt}
\newcommand{\bkt}{\beta \kt}
\newcommand{\ex}{{\rm e}}
\newcommand{\up}{$\uparrow$}
\newcommand{\down}{$\downarrow$}
\newcommand{\incr}{$\nearrow$}
\newcommand{\decr}{$\searrow$}
\newcommand{\onemin}{ \hspace*{-1.1mm}$\searrow\hspace{-2.5mm}
 {}_{{}_{{}_{\textstyle \smile}}}\hspace{-2.7mm}\nearrow$\hspace*{-1.1mm}}

\begin{document}


{\em {Institute for Condensed Matter Physics

National Academy of Sciences of Ukraine
        \footnote
         {) 1 Svientsitskii St., L'viv-11, 79011, Ukraine.

     \hspace{4mm}
     Tel: (0322) 707439

     \hspace{4mm}
     Fax: (0322) 761978

     \hspace{4mm}
         E-mail: ostb@icmp.lviv.ua})
}}

\vspace*{5mm}
{\bf
{Pair correlation functions of the Ising type model with spin 1
within two-particle cluster approximation
}}

\vspace*{4mm}

By

\vspace*{4mm}

O.R. BARAN, R.R. LEVITSKII

\vspace{5mm}

\noindent
The Blume-Emery-Griffiths model on hypercubic lattices
within the two-particle cluster approximation is investigated.
The expressions for the pair correlation functions in
$\bf{k}$-space are derived.
On the basis of obtained results (at $\bf{k}=0$)
the static susceptibility of
this model on the simple cubic lattice is calculated
at various values of the single-ion
anisotropy and biquadratic interaction.

\vspace{5mm}

\renewcommand\baselinestretch {1.5}
\large\normalsize

\section{Introduction}

The Blume-Emery-Griffiths (BEG) model
\begin{eqnarray}
\label{f1}
H= - \sum_{i=1}^N D_i S_i^2  -
\frac{1}{2} \sum_{i,\delta} \left[ K S_i S_{i+\delta} +
K' S_i^2 S_{i+\delta}^2 \right]  \;
\end{eqnarray}
(where $S_i=0,\pm 1$; $D_i$ is a single-ion anisotropy energy;
$K$ and $K'$ are the constants of bilinear and biquadratic
short-range interaction; the summation $i,\delta$ is going
over nearest neighbor pairs)
was originally proposed for the description of the phase
transition (PT) in He${}^3$ - He${}^4$ fluid \cite{beg1}.
This model has been extensively studied not only because of the
relative simplicity with which approximate calculations for the
model can be carried out and tested
as well as of the fundamental theoretical
interest arising from the richness of the phase diagram
that is exhibited due to competition of interactions,
but also because versions and extensions
of the model can be applied for the description of
simple and multi-component fluids \cite{beg2,beg3,beg4},
dipolar and quadrupolar orderings in magnets \cite{beg4,Chen,Nagaev},
crystals with ferromagnetic impurities \cite{beg4},
ordering in semiconducting alloys \cite{beg5}, etc.
The model has been studied within mean-field approximation (MFA)
\cite{beg1,beg2,beg3,beg4,Chen}, two-particle cluster approximation
(TPCA) \cite{tp1,my1,my2,my3}, effective field theory \cite{ef1,
ef2,ef3,ef4}, high-temperature series expansions \cite{ht1},
position--space renormalization group calculations \cite{rg1}, and
Monte Carlo simulations \cite{mc1,mc2,mc3,mc4}.

For compounds described by pseudospin models with essential short-range
correlations, the cluster approximation (CA) \cite{ca1,ca2,ca3,my2,my3}
is the most natural many-particle generalization of the MFA.
CA not only essentially improves the MFA results for the Ising
type model, but also is correct at those values of the
parameters, at which MFA gives qualitatively incorrect results.
Thus CA, in contrast to MFA, does not predict PT for the 1D
Ising model (with only bilinear short-range interaction) with an
arbitrary value of spin \cite{my1}, correctly responds to the
competition between antiferromagnetic biquadratic interaction
and ferromagnetic bilinear interaction in BEG model \cite{my2}.
Within CA, an infinite lattice is replaced with a cluster with a
fixed number of pseudospins; the influence of rejected sites is
taken into account as a single field $\varphi(S)$, acting on
boundary sites of a cluster.

The BEG model has a complicated phase diagram \cite{mc3,cc1}.
In \cite{cc1} the model was investigated within the
constant-coupling approximation (giving the same results as
TPCA) at $d=\frac{D}{K}<0$, $k'=\frac{K'}{K}>0$
for a cubic lattice. The phase diagram
$(\frac{D}{K'}, \frac{K}{K'}, T)$ and its projection onto the
$(\frac{D}{K'}, \frac{K}{K'})$ plane were constructed. The phase
transitions were classified. The authors presented the temperature
dependences of dipolar $m=\langle S \rangle$ and quadrupolar
$q=\langle S^2 \rangle$ moments for the some interesting sets of
parameters. In \cite{mc3} the BEG model on a simple
cubic lattice was studied within
Bethe approximation (giving the same results as TPCA)
and on the basis of the Monte Carlo simulation. It was shown
that for $d_a<d<-6k'-6$ at $k'<-1$ ($d_a$ depends on $k'$:
$d_a=0, -0.6, -1.2$ for $k'=-1, -2.25, -3.85$, respectively)
the phase with two sublattices $m_A$, $m_B$, $q_A$, $q_B$ exists.
The authors were particularly interested in the such sets of
model parameters where different kinds of re-entrant and double
re-entrant phase transitions took place.

The aim of the present paper is to calculate within TPCA
the pair correlation functions
$\langle (S_i)^l(S_j)^n \rangle^c =
\langle (S_i)^l(S_j)^n \rangle -
\langle (S_i)^l\rangle \cdot \langle(S_j)^n \rangle$
($l$ and $n=1,2$) of the BEG model in $\bf{k}$-space
and to investigate, using the obtained results (at $\bf{k}=0$)
the temperature dependences of static susceptibility of
the model on a simple cubic lattice at various values
of the single-ion anisotropy and biquadratic interaction
(in one-sublattice regions of the phase diagram, only).

\section{The two-particle cluster approximation}

The expression for a free enerqy within TPCA is constructed
on the basis of one-particle Hamiltonian $H_1$
\begin{eqnarray}
\label{f2}
H_1= - \kt_1 S_1 - \kt_1' S_1^2  ; \qquad
\kt_1 = \Gamma_1 +
\sum_{r \in {\textstyle \pi_1}} {}^r \! \varphi_1 ; \qquad
\kt_1' = D_1 + \sum_{r \in {\textstyle \pi_1}} {}^r \! \varphi_1'
\end{eqnarray}
(where the site $r$ is a nearest neighbour of the site $1$
($r \in \pi_1$)) and two-particle Hamiltonian $H_{12}$
\begin{eqnarray}
&& \label{f3}
H_{12}=- {}^2 \kt_1 S_1 - {}^1 \kt_2 S_2 - {}^2 \kt_1' S_1^2 -
{}^1 \kt_2' S_2^2 - K S_1 S_2 - K' S_1^2 S_2^2 ;
\\ && \nonumber
{}^1 \kt_2 = \Gamma_2 +
\sum_{ {\scriptstyle {}^{ r \in {\textstyle\pi_2}}_{ r \neq 1 }}}
{}^r \! \varphi_2 ; \qquad
{}^1 \kt_2' = D_2 +
\sum_{ {\scriptstyle {}^{ r \in {\textstyle\pi_2}}_{ r \neq 1 }}}
{}^r \! \varphi_2'
\end{eqnarray}
in a usual way \cite{my2,my3} (magnetic field $\Gamma_i \rightarrow0$
is introduced for convenience; $K>0$).
\begin{eqnarray}
\label{f4}
F = -k_B T \Big[(1-z) \sum_1 \ln \Big( {\rm Sp} \;\!
\ex^{-\beta H_1} \Big) + \frac 12
\sum_{1,2} \ln \Big( {\rm Sp} \;\! \ex^{-\beta H_{12}}\Big) \Big]
\end{eqnarray}
Here $z$ is the number of nearest neighbours and $\beta=1/(k_BT)$.
In the case when the fields are uniform, the free energy can
be written as:
\begin{eqnarray}
&& \label{f5}
F = -k_B T N \Big[ (1-z) \ln Z_1 + \frac{z}{2} \ln Z_{12} \Big] ;
\\ && \label{f6}
Z_1 = 2 \ex^{ \bkt'} \cdot \cosh( \bkt ) +1 ;
\\ && \nonumber
Z_{12} = 2 \ex^{ \beta ( 2 \ktt' + K' )}
\Big( \ex^{ \beta K} \cdot \cosh(2 \bktt ) +
\ex^{-\beta K} \Big) + 4 \ex^{ \bktt' } \cdot \cosh( \bktt ) + 1 ,
\end{eqnarray}
where
\[
\kt = \Gamma + z \varphi ; \quad
\kt' = D + z \varphi'; \quad
\ktt = \Gamma + (z-1) \varphi ; \quad
\ktt' = D + (z-1) \varphi'.
\]

The cluster parameters $\varphi$ and $\varphi'$ are found by
minimizing the free energy with respect to them. The following
system of equation for $\varphi$ and $\varphi'$ is obtained:
\begin{eqnarray}
&& \label{f7} \hspace*{-8mm}
\frac{ \ex^{ \bkt'} \cdot \sinh( \bkt)}{Z_1} =
\frac{ \ex^{ \beta (2\ktt'+K'+K)} \cdot \sinh(2 \bktt)
+ \ex^{ \bktt'} \cdot \sinh( \bktt)  }{ Z_{12} }  ;
\\ && \nonumber \hspace*{-8mm}
\frac{ \ex^{ \bkt'} \cdot  \cosh( \bkt)}{ Z_1 } =
\frac{ \ex^{ \beta (2\ktt'+K')} \left[ \ex^{ \beta K} \cdot
\cosh(2 \bktt) + \ex^{- \beta K} \right] + \ex^{ \bktt'} \cdot
\cosh( \bktt) } { Z_{12} }.
\end{eqnarray}
Using (\ref{f7}) we can write simple expressions for
magnetization $m=\ang{S}$ and quadrupolar moment $q=\ang{S^2}$:
\begin{eqnarray}
\label{f8}
m = \frac{2 \ex^{ \bkt'} \cdot \sinh( \bkt)}{Z_1} ; \qquad
q = \frac{2 \ex^{ \bkt'} \cdot \cosh( \bkt)}{Z_1} .
\end{eqnarray}

The correlation functions can be found by differentiating the
free enerqy (\ref{f4}) of the system in nonuniform external
fields ($\Gamma_i$, $D_i$) with respect to these fields.
In the case of a $\Delta$-dimensional hypercubic
lattice the matrix of pair correlation functions
(in the uniform fields case)
in $\bf{k}$-space has the form \cite{my2,my3}:
\begin{eqnarray}
\label{f9}
\left(\begin{array}{ll}
\ang{S_{\bf k} S_{-{\bf k}}}^c &
\ang{S_{\bf k} S^2_{-{\bf k}}}^c \\
\ang{S^2_{\bf k} S_{-{\bf k}}}^c &
\ang{S^2_{\bf k} S^2_{-{\bf k}}}^c
\end{array}\right)
&=& \bigg[ (1-z)\Big(\hat{G}_{1}\Big)^{-1} +
z\Big(\hat{G}_{1}+\hat{G}_{12}\Big)^{-1}
\\ \nonumber &+& 4\Big[\hat{G}_{1}
\Big(\hat{G}_{12}\Big)^{-1}\hat{G}_{1}-\hat{G}_{12}\Big]^{-1}
\sum^{\Delta}_{a=1} \sin^2(\frac{k_a}{2}) \Bigg]^{-1}\;
\end{eqnarray}
(let us note that
$\ang{S_{\bf k} S^2_{-{\bf k}}}^c =
\ang{S^2_{\bf k} S_{-{\bf k}}}^c $).
Here the following matrices of one-particle and two-particle
intracluster correlation functions are introduced:
\begin{eqnarray}
\label{f10}
\hat{G}_{1}=
\left(\begin{array}{ll}
\ang{S_1 S_1}^c_{H_1} & \ang{S_1 S_1^2}^c_{H_1} \\
\ang{S_1^2 S_1}^c_{H_1} & \ang{S_1^2 S_1^2}^c_{H_1}
\end{array}\right), \quad
\hat{G}_{12}=
\left(\begin{array}{ll}
\ang{S_1 S_2}^c_{H_{12}} & \ang{S_1 S_2^2}^c_{H_{12}} \\
\ang{S_1^2 S_2}^c_{H_{12}} & \ang{S_1^2 S_2^2}^c_{H_{12}}
\end{array}\right) .
\end{eqnarray}

\section{Numerical analysis results}

In this section we discuss the results of numerical calculations
within TPCA for temperature dependences of static susceptibility
$\chi=\beta \ang{S_{\bf{k}} S_{-\bf{k}}}^c_{|_{\bf{k}=0}}$
of BEG model on a simple cubic lattice ($z=6$).

Here we use the following notations
for the relative quantities:
$t=(3k_B T)/(2 z K)$, $d=D/K$, $k'=K'/K$;
and the terminology of \cite{cc1}:
{\bf F} -- the ferromagnetic phase
($m \ne 0$, $q \ne {\textstyle \frac{2}{3}}$),
{\bf P} -- the paramagnetic phase
($m = 0$, $q \ne {\textstyle \frac{2}{3}}$,
$q(t\rightarrow\infty)={\textstyle \frac{2}{3}}$),
{\bf Q} -- the quadrupolar phase
($m = 0$, $q \ne {\textstyle \frac{2}{3}}$).
In the two-particle cluster approximation the system of equations for
$\varphi$, $\varphi'$ (\ref{f7}) has several solutions, the number of
which depends on values of parameters $d$, $k'$ and temperature.
Solution corresponding to the {\bf P} phase exists at
$t \in [t_{{\scriptscriptstyle {\rm P}}_1},
\infty]$ ($t_{{\scriptscriptstyle {\rm P}}_1} \geq 0$, its value
depends on $d$, $k'$).
Solutions corresponding to the
{\bf F} phase and {\bf Q} phase exist at
$t \in [t_{{\scriptscriptstyle {\rm F}}_1},
t_{{\scriptscriptstyle {\rm F}}_2}]$ and
$t \in [t_{{\scriptscriptstyle {\rm Q}}_1},
t_{{\scriptscriptstyle {\rm Q}}_2}]$,
respectively. The values of
$t_{{\scriptscriptstyle {\rm F}}_1}$,
$t_{{\scriptscriptstyle {\rm F}}_2}$ and
$t_{{\scriptscriptstyle {\rm Q}}_1}$,
$t_{{\scriptscriptstyle {\rm Q}}_2}$
depend on $d$, $k'$ and are finite.

The projection of the phase diagram on ($d,k'$) plane for
ferromagnetic bilinear interaction at $d<0$, $k'>0$ \cite{cc1}
and $d\geq0$, $k'>-1-\frac{1}{6} d$ (see fig. 1) consists of
seven regions:
I -- the first order phase transition
{\bf Q} $\leftrightarrow$ {\bf P} ({\bf QP}1),
II -- the PT {\bf FP}2,
III -- the PT {\bf FP}1,
IV -- the PT is absent (the system is in the {\bf P} phase),
V -- the PTs {\bf QF}1 and {\bf FP}2,
VI -- the PTs {\bf QF}1 and {\bf FP}1,
VII -- the PTs {\bf FQ}1 and {\bf QP}1.

Let us consider now the temperature dependence of the inverse
static susceptibility $\chi^{-1}(t)$
along with quadrupolar moment $q(t)$
(the latter has been already studied in Ref. \cite{cc1}).
In the {\bf F} phase $q(t)$ and $\chi^{-1}(t)$ decrease
as $t$ increases (see figs. 2-6). In the {\bf Q} phase $q(t)$
increases and $\chi^{-1}(t)$ decreases (see figs. 5-7).
In the {\bf P} phase the situation is more complicated.
Depending on the model parameters,
$q(t)$ can decrease or increase, and
$\chi^{-1}(t)$ can be an increasing function (see figs. 2, 6, 7)
or a non-monotonic function with one minimum (see figs. 3-5).
At infinitely high temperature
$\chi(t\rightarrow\infty)=2/3 \cdot \beta$.

It should be noted that at any ferromagnetic set of the model
parameters ($d\geq0$, $k'\geq0$), the quadrupolar moment
(in the {\bf P} phase) is lowered down by $t$,
and at ($d$, $k'$) from region IV it is raised up.
The fact that decreasing behavior of $q(t)$
in the {\bf P} phase is changed to an increasing one
is caused by decreasing of $d$ or $k'$.
At the set of the model parameters from regions II or V,
$\chi^{-1}(t)$ in the {\bf P} phase is an increasing function,
and at ($d$, $k'$) from region IV it is a non-monotonic function.
Non-monotonic behavior of $\chi^{-1}(t)$ in the {\bf P} phase
is possible only at those $d$ and $k'$,
at which $q(t)$ increases in the {\bf P} phase.
That is, for instance, increasing of antiferromagmetic $d$
at constant ferromagnetic $k'$ must give rise, first, to
increasing of $q(t)$, and only then to
non-monotonic behavior of $\chi^{-1}(t)$.

At those sets of the model parameters when the phase transition
{\bf FP}2 takes place in the system as $t$ increases
(region II and V; see fig. 6) $q(t_c)$ and $\chi^{-1}(t_c)$
have cusps ($\chi^{-1}(t_c)=0$) at the transition point.
In the {\bf P} phase, as has been already mentioned,
$\chi^{-1}(t)$ can only increase, and $q(t)$ can either decrease
or increase. Increasing of $q(t)$ is possible only in a small
part of region II at $-3<k'<1.5$ and sufficiently
small $d$ (near the regions III or $d\geq0$, $k'<-1-\frac 16 d$).

At those sets of the model parameters when the phase
transition {\bf FP}1 takes place in the system
(regions III and VI; see figs. 2-4) $q(t_c)$ and
$\chi^{-1}(t_c)$ have finite jumps at the transition point,
and only $q(t_c)$ always has a downward one
($q(t_c-0)>q(t_c+0)$).
All possible combinations of the $\chi^{-1}(t_c)$ jump and
behaviors of $\chi^{-1}(t)$ and $q(t)$ in the {\bf P} phase,
depending on the model parameters, are the following:
\begin{table}[h]
\centerline{ \begin{tabular}{|l|c|c|c|c|}    \hline
 &1 &2 &3 &4  \\   \hline
$q(t)$ &\decr &\incr &\incr &\incr \\   \hline
$\chi^{-1}(t)$ &\incr &\incr &\onemin &\onemin \\ \hline
$\chi^{-1}(t_c)$ &\down &\down &\down &\up \\ \hline
\end{tabular} }
\end{table} \\
(hereafter we use the following notations:
\incr (\decr) -- increasing (decreasing) function,
\onemin \hspace*{1.1mm} -- non-monotonic function with one minimum,
\up (\down) -- function has a finite upward (downward) jump).
The combinations 1,3,4 are presented in figs.  2-4, respectively.
It should be noted that an upward jump of $\chi^{-1}(t_c)$
(which can take place only in a small part of region III)
is possible when $\chi^{-1}(t)$ is non-monotonic
function in the {\bf P} phase, only.
Thus, increasing of antiferromagnetic $d$
(or decreasing of ferromagnetic $k'$), first must induce a
non-monotonic behavior of $\chi^{-1}(t)$, and only then an
upward jump of $\chi^{-1}(t_c)$.
But it is possible (depending on $k'$)
that further increasing of antiferromagnetic $d$ leads to a
downward jump of $\chi^{-1}(t_c)$ again.
For instance, the sequences of presented combinations of
$\chi^{-1}(t_c)$ jump and behaviors of $q(t)$ and $\chi^{-1}(t)$
in the {\bf P} phase as antiferromagnetic $d$ increases
(at given $k'$) are: 2, 3 at $k'=0.0$;
1, 2, 3, 4, 3 at $k'=2.0$; 1, 2, 3, 4 at $k'=2.6$;
1, 2 at $k'=2.95$; at $k'=3.2$ the combination 1 is possible only
(at $k'=0.0$ $q(t)$ becomes increasing yet in the region II).
It should also be noted that an upward jump of $\chi^{-1}(t_c)$
and concavity of the $q(t)$ curve in the {\bf P} phase at low
temperatures (as in fig. 4), are independent phenomena.

At those sets of the model parameters, when the phase transition
{\bf QP}1 takes place in the system (regions I and VII;
see figs. 5, 7) at the transition point $q(t_c)$ and
$\chi^{-1}(t_c)$ have finite upward and downward
jumps, respectively. In the {\bf P} phase, behaviors of
$q(t)$ and $\chi^{-1}(t)$ can be the following:
\begin{table}[h]
\centerline{ \begin{tabular}{|l|c|c|c|}    \hline
 &1 &2 &3  \\   \hline
$q(t)$ &\decr &\incr &\incr \\   \hline
$\chi^{-1}(t)$ &\incr &\incr &\onemin  \\ \hline
\end{tabular} }
\end{table} \\
(the first and third combinations are presented in figs. 7
and 5, respectively).
It should be remembered that the order of presented combinations
of quadrupolar moment and inverse static susceptibility
temperature behaviors in the {\bf P} phase
corresponds to increasing of antiferromagnetic $d$
(or decreasing of ferromagnetic $k'$).
For instance, as antiferromagnetic $d$ increases, the sequence
of presented combinations is 2, 3 at $k'=2.95$
and 1, 2, 3 at $k'=4.0$
(at $k'=2.95$ $q(t)$ becomes increasing yet in the region III).

At the {\bf FQ}1 phase transition (region VII ; see fig. 5)
$q(t_c)$ and $\chi^{-1}(t_c)$ have finite
downward and upward jumps, respectively.
At the {\bf QF}1 transition (regions V, VI; see fig. 6)
the situation is reverse ($q(t_c)$ has an upward jump,
and $\chi^{-1}(t_c)$ has a downward one).

Let us note that the proection of the phase diagram on ($d, k'$)
plane at $d<0$, $k'<0$ and $d \geq 0$, $k'<-1-\frac{1}{6}d$
is also complicated \cite{mc3}, and study of static susceptibility at
those sets of model parameters is subject of a separate paper.

\section{Conclusions}
In the present paper, the pair correlation functions in
$\bf{k}$-space for the Blume-Emery-Griffiths model have
been obtained.

The temperature dependences of static susceptibility
at various values of model parameters have been
investigated. It was shown that in the paramagnetic phase it
can be a non-monotonic function of temperature with one maximum.
Such behavior is impossible at those sets of the model parameters,
when the second order phase transition ferromagnet -- paramagnet
takes place, and becomes possible after the first order phase
transitions from the ferromagnetic or quadrupolar phases
(only when the quadrupolar moment in the paramagnetic phase
is an increasing function).

It should be noted that if we know the temperature dependences of
quadrupolar moment and static susceptibility only in a narrow
temperature interval, and those quantities increase (furthermore,
the curve of quadrupolar moment temperature dependence is concave),
that does not suffice to state that the system is in a
quadrupolar, not in a paramagnetic phase. However, either
decreasing of static susceptibility, or convexity (concavity) of the
increasing (decreasing) quadrupolar moment curve is enough to state
that the system is in the paramagnetic, not in the quadrupolar phase.

\pagebreak

\begin{figure}[h]
\begin{center}
\leavevmode
\mframe{\epsffile{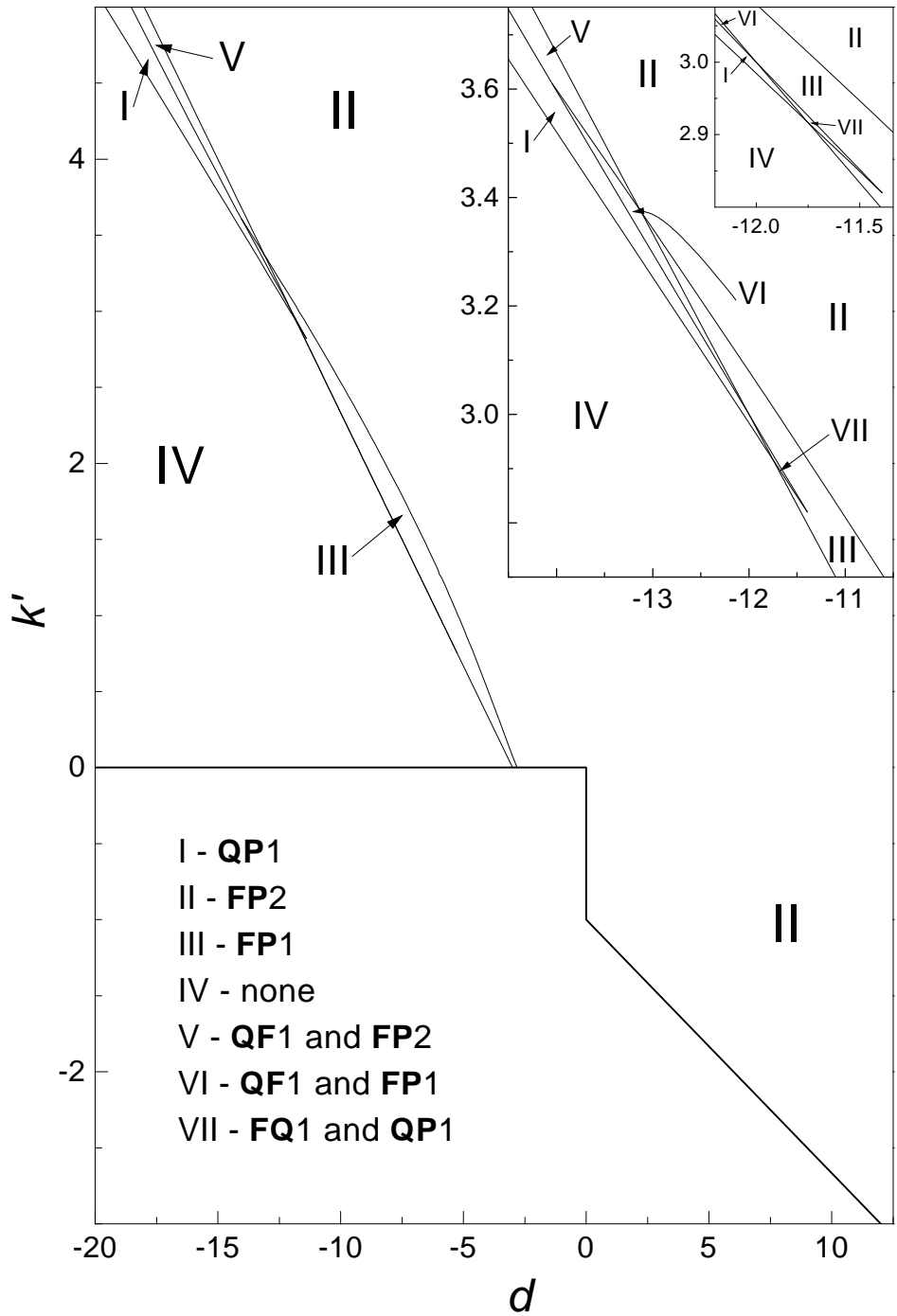}}
\end{center}
\vspace{-6mm} \caption{\small The proection of the phase diagram
onto $(d, k')$ plane. } \label{fig1}
\end{figure}

\clearpage
\begin{figure}[h]
\begin{center}
\leavevmode
\mframe{\epsffile{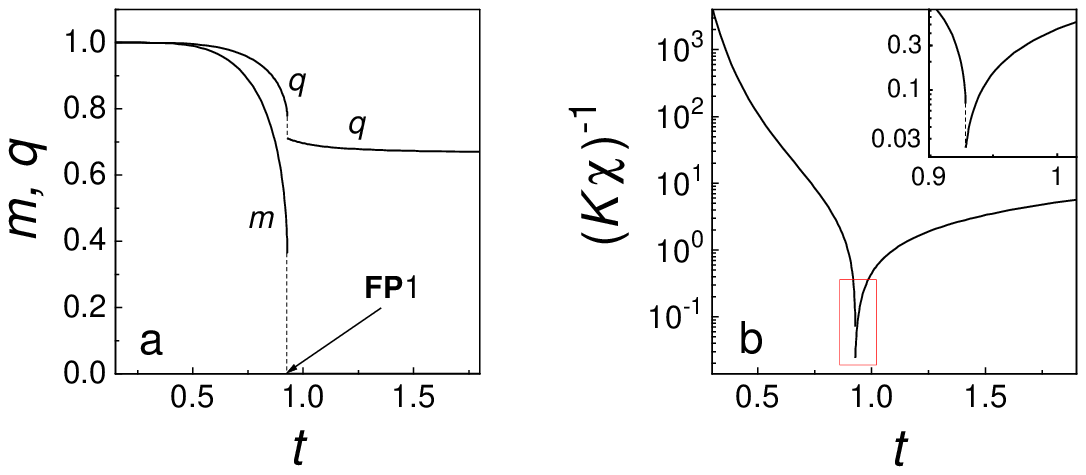}}
\end{center}
\vspace{-5mm} \caption{\small The temperature dependences of $m$,
$q$ and inverse static susceptibility $\chi^{-1}$ at $k'=2.6$,
$d=-10.4$. } \label{fig2}
\vspace*{5mm}
\begin{center}
\leavevmode
\mframe{\epsffile{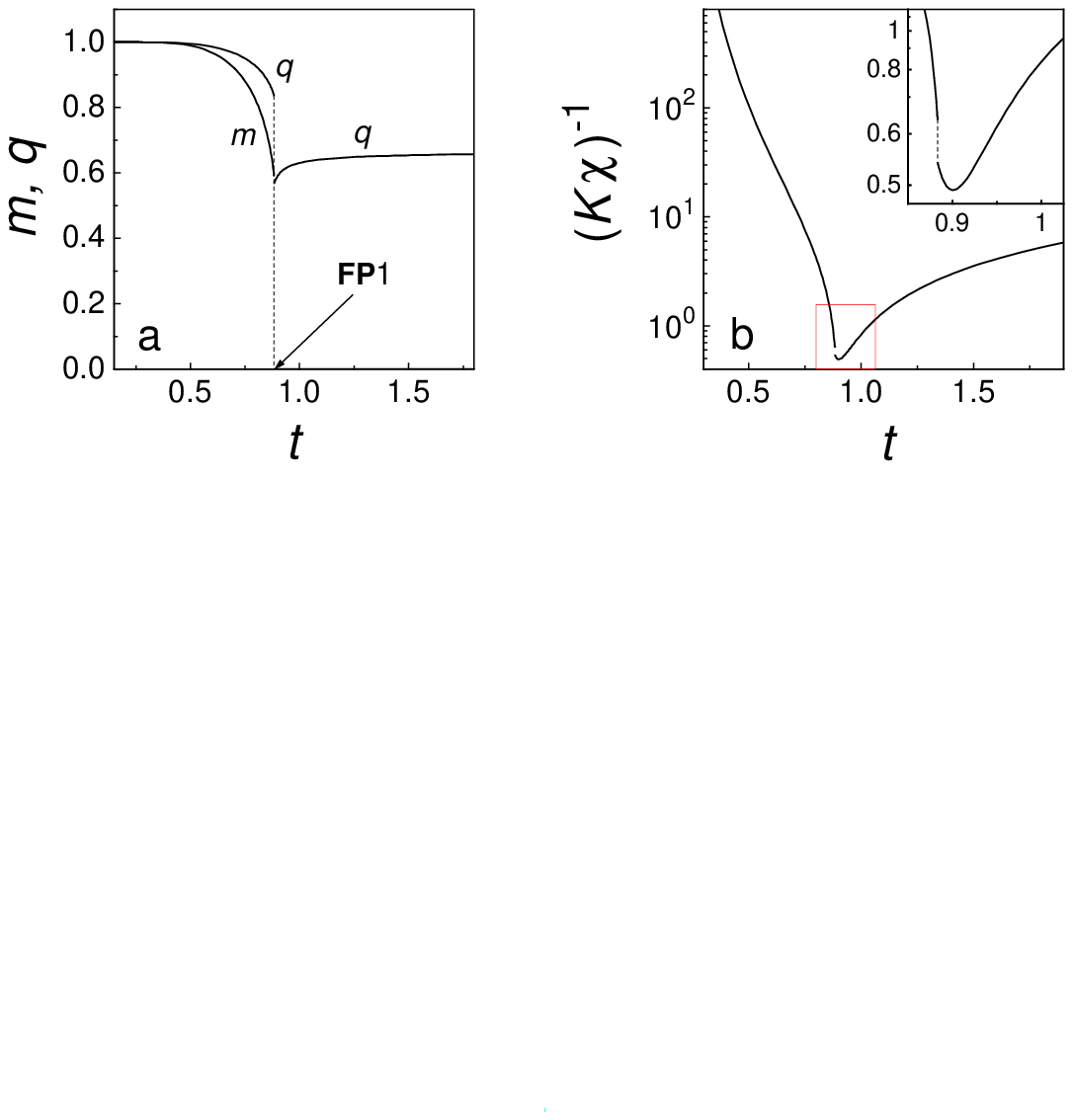}}
\end{center}
\vspace{-5mm} \caption{\small The temperature dependences of $m$,
$q$ and inverse static susceptibility $\chi^{-1}$ at $k'=2.6$,
$d=-10.63$. } \label{fig3}
\end{figure}

\clearpage
\begin{figure}[h]
\begin{center}
\leavevmode
\mframe{\epsffile{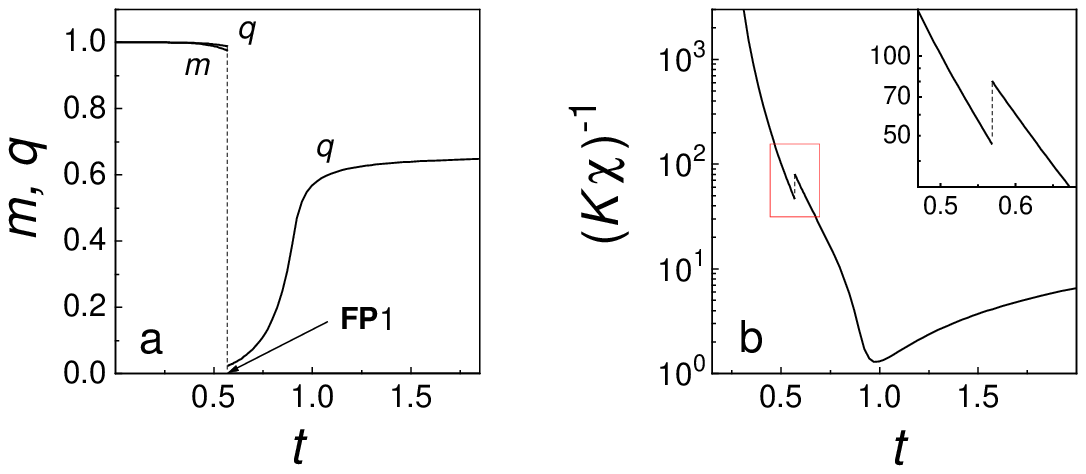}}
\end{center}
\vspace{-5mm} \caption{\small The temperature dependences of $m$,
$q$ and inverse static susceptibility $\chi^{-1}$ at $k'=2.6$,
$d=-10.79$. } \label{fig4}
\vspace*{5mm}
\begin{center}
\leavevmode
\mframe{\epsffile{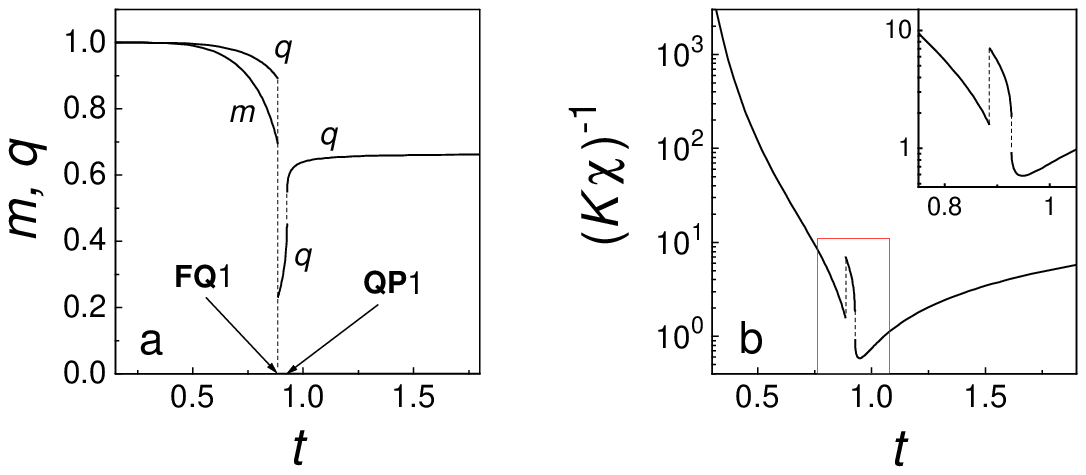}}
\end{center}
\vspace{-5mm} \caption{\small The temperature dependences of $m$,
$q$ and inverse static susceptibility $\chi^{-1}$ at $k'=2.88$,
$d=-11.61$. } \label{fig5}
\end{figure}

\clearpage
\begin{figure}[h]
\begin{center}
\leavevmode
\mframe{\epsffile{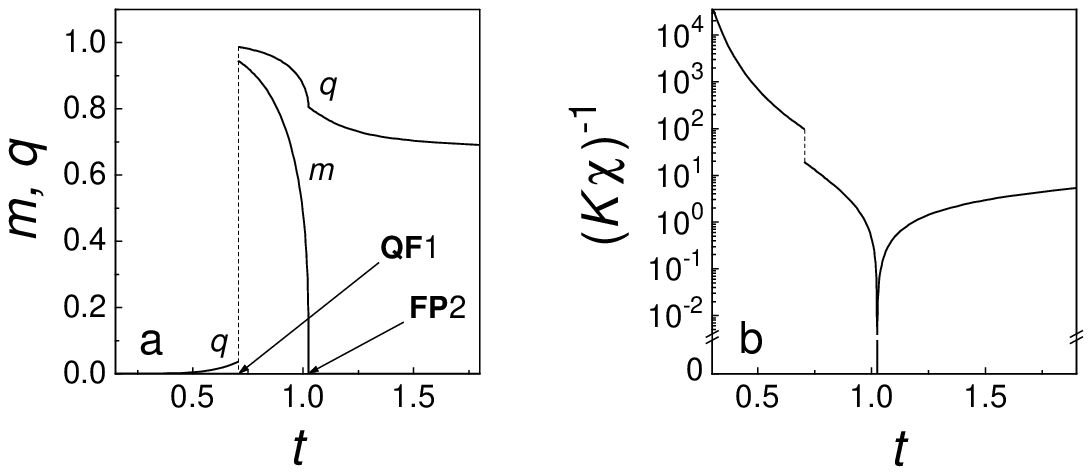}}
\end{center}
\vspace{-5mm} \caption{\small The temperature dependences of $m$,
$q$ and inverse static susceptibility $\chi^{-1}$ at $k'=3.44$,
$d=-13.34$. } \label{fig6}
\vspace*{5mm}
\begin{center}
\leavevmode
\mframe{\epsffile{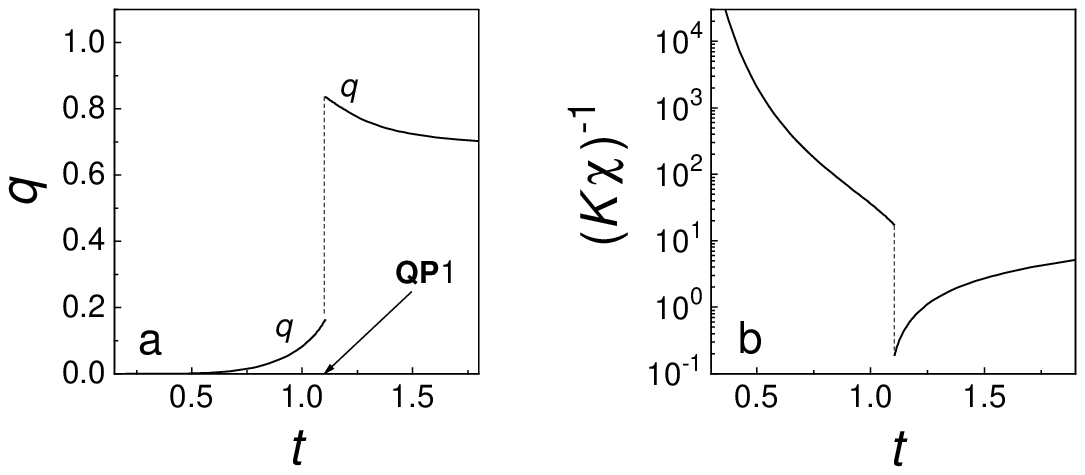}}
\end{center}
\vspace{-5mm} \caption{\small The temperature dependences of $q$
and inverse static susceptibility $\chi^{-1}$ at $k'=4.0$,
$d=-15.4$. } \label{fig7}
\end{figure}

\end{document}